\documentclass[%
reprint,
superscriptaddress,
amsmath,amssymb,
aps,
pra]{revtex4-2}

\usepackage{textgreek}
\usepackage{nicefrac}
\usepackage{siunitx}
\usepackage{multirow}
\usepackage[normalem]{ulem}
\usepackage{graphicx}
\usepackage{dcolumn}
\usepackage{bm}
\usepackage{microtype}
\usepackage[final]{changes}

\usepackage[
   colorlinks=true,
   citecolor=blue,
   linkcolor=blue,
   urlcolor=blue,
]{hyperref}
\usepackage{orcidlink}

\newcommand{\NxLO}[2][]{
    \ifcase#2\ensuremath{\text{LO}^{#1}}
    \or\ensuremath{\text{NLO}^{#1}}
    \else\ensuremath{\text{N}^{#2}\text{LO}^{#1}}
    \fi
}

\newcommand{\nlo}{\ensuremath{\mathrm{NLO}}}
\newcommand{\nnlo}{\ensuremath{\mathrm{N}^2\mathrm{LO}}}
\newcommand{\nnnlo}{\ensuremath{\mathrm{N}^3\mathrm{LO}}}

\begin{document}

\title{Accurate Charge Radius Measurement of $^{14}$C Confronts \textit{Ab Initio} Theory}

\author{Kristian König\orcidlink{0000-0001-9415-3208}}
\email{kkoenig@ikp.tu-darmstadt.de}
\affiliation{Institut f\"ur Kernphysik, Technische Universit\"at Darmstadt, 64289 Darmstadt, Germany}%
\affiliation{Helmholtz Research Academy Hesse for FAIR, GSI Darmstadt, 64291 Darmstadt, Germany}

\author{Patrick M\"{u}ller\orcidlink{0000-0002-4050-1366}}
\affiliation{Institut f\"ur Kernphysik, Technische Universit\"at Darmstadt, 64289 Darmstadt, Germany}

\author{Tobias Gesser\orcidlink{0009-0006-5401-7045}}
\affiliation{Institut f\"ur Kernphysik, Technische Universit\"at Darmstadt, 64289 Darmstadt, Germany}%

\author{Emily Burbach\orcidlink{0009-0001-4237-0718}}
\affiliation{Institut f\"ur Kernphysik, Technische Universit\"at Darmstadt, 64289 Darmstadt, Germany}

\author{Stefano Gandolfi\orcidlink{0000-0002-0430-9035}}
\affiliation{Theoretical Division, Los Alamos National Laboratory, Los Alamos, New Mexico 87545, USA}%

\author{Matthias Heinz\orcidlink{0000-0002-6363-0056}}
\affiliation{National Center for Computational Sciences, Oak Ridge National Laboratory, Oak Ridge, TN 37831, USA}
\affiliation{Physics Division, Oak Ridge National Laboratory, Oak Ridge, TN 37831, USA}

\author{Phillip Imgram\orcidlink{0000-0002-3559-7092}}
\affiliation{Institut f\"ur Kernphysik, Technische Universit\"at Darmstadt, 64289 Darmstadt, Germany}

\author{Alessandro Lovato \orcidlink{0000-0002-2194-4954}}
\affiliation{Physics Division, Argonne National Laboratory, Argonne, Illinois 60439, USA}
\affiliation{Computational Science Division, Argonne National Laboratory, Argonne, Illinois 60439, USA}
\affiliation{INFN-TIFPA Trento Institute of Fundamental Physics and Applications, I-38123 Povo, Trento, Italy}
\affiliation{Instituto de Física Corpuscular (IFIC), Consejo Superior de Investigaciones Científicas (CSIC) and Universidad de Valencia
E-46980 Paterna, Valencia, Spain}

\author{Pieter Maris\orcidlink{0000-0002-1351-7098}}
\affiliation{Dept. of Physics and Astronomy, Iowa State University, Ames, Iowa 50011, USA}

\author{Takayuki Miyagi\orcidlink{0000-0002-6529-4164}}
\affiliation{Center for Computational Sciences, University of Tsukuba, 1-1-1 Tennodai, Tsukuba 305-8577, Japan}

\author{Wilfried N\"ortersh\"auser\orcidlink{0000-0001-7432-3687}}
\affiliation{Institut f\"ur Kernphysik, Technische Universit\"at Darmstadt, 64289 Darmstadt, Germany}%
\affiliation{Helmholtz Research Academy Hesse for FAIR, GSI Darmstadt, 64291 Darmstadt, Germany}

\author{Robert Roth\orcidlink{0000-0003-4991-712X}}
\affiliation{Institut f\"ur Kernphysik, Technische Universit\"at Darmstadt, 64289 Darmstadt, Germany}%
\affiliation{Helmholtz Research Academy Hesse for FAIR, GSI Darmstadt, 64291 Darmstadt, Germany}

\author{Julien Spahn\orcidlink{0009-0007-8354-4896}}
\affiliation{Institut f\"ur Kernphysik, Technische Universit\"at Darmstadt, 64289 Darmstadt, Germany}%

\author{Achim Schwenk\orcidlink{0000-0001-8027-4076}}
\affiliation{Institut f\"ur Kernphysik, Technische Universit\"at Darmstadt, 64289 Darmstadt, Germany}
\affiliation{ExtreMe Matter Institute EMMI, GSI Helmholtzzentrum f\"ur Schwerionenforschung GmbH, 64291 Darmstadt, Germany}
\affiliation{Max-Planck-Institut f\"ur Kernphysik, 69117 Heidelberg, Germany}

\begin{abstract}
Located at the neutron shell closure $N = 8$, the long-lived radioactive isotope \(^{14}\mathrm{C} \) plays a critical role in geochronology and nuclear structure studies. Despite its widespread use, the nuclear charge radius of $^{14}$C has remained less precisely known compared to its stable counterpart $^{12}$C. Here, we report a high-precision determination of the $^{14}$C charge radius using collinear laser spectroscopy at the COALA setup at TU Darmstadt, improving upon the precision of previous muonic measurements by a factor of $5$ and revealing a $1.9\sigma$ discrepancy of combined uncertainty, indicating a likely underestimated uncertainty in the muonic determination.
This measurement challenges state-of-the-art \textit{ab initio} nuclear theory calculations, including auxiliary field diffusion Monte Carlo, the valence-space in-medium similarity renormalization group, and the no-core shell model, augmented by neural-network techniques. 
With $^{12}$C and $^{14}$C now forming one of the most precisely characterized even-even isotope pairs, these results also enable improved QED tests.
\end{abstract}

\maketitle

{\it Introduction.}
Located at the $N=8$ neutron shell closure, $^{14}$C is the lightest neutron-rich radioactive carbon isotope. 
Its long lifetime of 5700 years makes it ideal for geological and archaeological age dating~\cite{Kutschera.2019} and macroscopic samples become available, e.g., from neutron capture at nuclear reactors.
This enables precision measurements of its fundamental nuclear properties like its mass, precisely measured with RF mass spectrometry~\cite{Smith.1975}, and its charge radius, investigated by elastic electron scattering~\cite{Kline.1973} and muonic atom x-ray spectroscopy~\cite{Schaller.1982}. The extracted charge radii are significantly less precise than for the stable $^{12}$C~\cite{Sick.1982,Ruckstuhl.1984} but still considerably more precise than for the more neutron-rich isotopes extracted from charge-changing cross section measurements~\cite{Yamaguchi.2011, Kanungo.2016, Zhao.2024}. 

The carbon isotopic chain is rich in structure: Proposed are halo effects in $^{15,16,19,22}$C~\cite{Kanungo.2016,Fang.2004, Horiuchi.2006, Tanaka.2010, Mouadil.2024} and antihalo characteristics in $^{16}$C~\cite{Matsumoto.2014}, along with signatures of magicity at the $N=16$ subshell closure~\cite{Ozawa.2000}.
The intrinsic structure of carbon isotopes is also of wide interest due to the tendency to form $\alpha$ clusters. In $^{12}$C, the $0_2^+$ Hoyle state~\cite{Hoyle.1954, Freer.2014} is essential in the 3$\alpha$-nucleosynthesis process in stars and supernovae~\cite{Fynbo.2005, Jin.2020}. In heavier isotopes, e.g., in $^{14}$C, the additional neutrons can be treated as additional clusters~\cite{Grinyuk.2023} and form linear~\cite{Kahl.2023, Han.2023} or even triangular configurations~\cite{Itagaki.2004}.

Here, we present a fivefold improvement in the ground-state charge radius of $^{14}\mathrm{C}$ obtained from collinear laser spectroscopy 
performed at the \textbf{Co}llinear \textbf{A}pparatus for \textbf{L}aser spectroscopy and \textbf{A}pplied physics (COALA) at TU Darmstadt~\cite{Konig.2020}.
The COALA setup has been used for a range of high-precision laser spectroscopy studies in stable isotopes, e.g., in Ba$^+$ and Ca$^+$~\cite{Imgram.2019,Muller.2020}, allowed for an all-optical determination of the $^{12}$C charge radius~\cite{Imgram.2023} and a more precise determination of the $^{13}$C charge radius~\cite{Muller.2025}.
We have now adapted the apparatus to handle moderately radioactive samples and used it to measure the $^{14}$C charge radius.
Now $^{12,14}$C is the second most precisely known pair of nuclear charge radii after $^{4,6}$He~\cite{Wang.2004,Mueller.2007}. This allows, e.g., for precise QED tests from $g$-factor measurements~\cite{Sailer.2022} or tests of state-of-the-art nuclear \textit{ab initio} theories~\cite{Miyagi:2025lmv}, for which we performed new calculations using the no-core shell model (NCSM) \cite{Barrett:2013nh} combined with artificial neural networks~\cite{WoKno24,WoGe25}, auxiliary field diffusion Monte Carlo (AFDMC)~\cite{Schmidt:1999lik,Carlson:2014vla, Gandolfi:2020pbj}, and the valence-space in-medium similarity renormalization group (VS-IMSRG)~\cite{Hergert:2015awm, Stroberg:2019mxo}.
Improving our knowledge of the charge radius is crucial for studies of mirror partners, including $^{14}$C--$^{14}$O, which are used to provide information about the size of the neutron skin~\cite{Yang.2018,Novario.2023}, the nuclear-matter equation of state~\cite{Brown.2020,Pineda.2021}, and the mass-radius relation of neutron stars~\cite{Bano.2023}. 
The successful demonstration of spectroscopy in the radioactive $^{14}$C paves the way for further measurements of short-lived carbon isotopes at accelerator facilities like ISOLDE as well as for the investigation of other long-lived nuclei at COALA.

{\it Experiment.}
In first order, differential mean-square nuclear charge radii $\delta \langle r_\mathrm{c}^2 \rangle^{A,A'}=\langle r_\mathrm{c}^2 \rangle^{A}-\langle r_\mathrm{c}^2 \rangle^{A'}$ can be determined from the isotope shift $\delta \nu^{A,A'}$, measured as the difference in transition frequency of two isotopes with mass numbers $A$ and $A'$,
\begin{equation}
\delta \nu^{A,A'}=\nu^A-\nu^{A'} \approx \delta\nu_\mathrm{M} + F \delta \left\langle r_\mathrm{c}^2 \right\rangle ^{A,A'},
\label{eq:r2}
\end{equation} 
if the mass-shift contribution $\delta\nu_\mathrm{M}$ and the field-shift parameter $F$ are known.
However, neither carbon atoms nor ions have laser-accessible transitions from the ground state. Populating a metastable neutral state, e.g., via electron transfer in a charge-exchange cell~\cite{Vernon.2019}, is in principle possible, but accurate atomic calculations for $\delta\nu_\mathrm{M}$ and $F$ are lacking. In particular, mass-shift calculations including electron correlations are notoriously difficult and have so far been achieved only for up to five-electron systems using NRQED~\cite{Maass.2019}.
For these reasons, we performed spectroscopy from the ortho-helium-like $1s2s\,^3\!S_1$ level ($2\,411\,292.8$\,cm$^{-1}$) in C$^{4+}$, which has a lifetime of $21$\,ms and is efficiently populated inside an Electron Beam Ion Source (EBIS)~\cite{ImgramPRA.2023}. From there, the $1s2p\,^3\!P_J$ fine-structure triplet is accessible at a wavelength of about $227$\,nm. 
The produced ions are electrostatically accelerated out of the EBIS, which effectively compresses the large velocity distribution from the hot EBIS plasma~\cite{Kaufman.1976}, and the fast beam is transported within a few $10\,\mu$s  to the detection region. 
The COALA setup at TU Darmstadt employed for these measurements is described in detail in~\cite{Konig.2020,ImgramPRA.2023} and here we give only a short overview. 

To produce a $^{14}$C$^{4+}$ beam, an EBIS-A (DREEBIT GmbH) was operated with an electron beam current of \SI{80}{\milli\ampere} and at a gas pressure of $6\cdot 10^{-8}$\,mbar of CO$_2$ gas, which was composed of $50$\,\% $^{14}$C and $50$\,\% $^{12}$C. The central trap electrode was floated at $12.5$\,kV, resulting in a beam energy of $50$\,keV. The EBIS was operated in the continuous beam mode, i.e., with a low trap exit electrode voltage that allows the ions to leak out in longitudinal direction, and tuned to generate a large fraction of C$^{4+}$ ions. 
A beam current of $0.8$\,nA of $^{14}$C$^{4+}$ ions was achieved at an exit trap potential of about $200$\,V above the central electrode potential~\cite{ImgramPRA.2023}.
A $60^\circ$ electrostatic bender guided the ion beam into the central COALA beamline, which has laser access from collinear and anticollinear direction. Ion optical steering and focusing elements were used to superpose the ion and laser beams. The beam overlap was checked at two diagnostic stations that are placed at a $2.6$-m distance and equipped with movable iris diaphragms, Faraday cups, and phosphor screens. The laser-ion interaction takes place in a fluorescence detection region, which houses a mirror-based and a lens-based light collection system to guide the fluorescence light to photomultipliers~\cite{Mueller.2024}. The fluorescence detection region is installed between the diagnostic stations and can be floated on an additional potential. For the present measurements a voltage of about $-200$\,V was applied, and the laser frequencies were adjusted such that the Doppler-shifted resonance condition is only met in the fluorescence detection region. Instead of varying the laser frequency to scan across the resonance, this potential is scanned by $10$\,V, which corresponds to a frequency variation of $1.5$\,GHz in the ions' rest frame.
\begin{figure}
    \centering
    \includegraphics[width=1\linewidth]{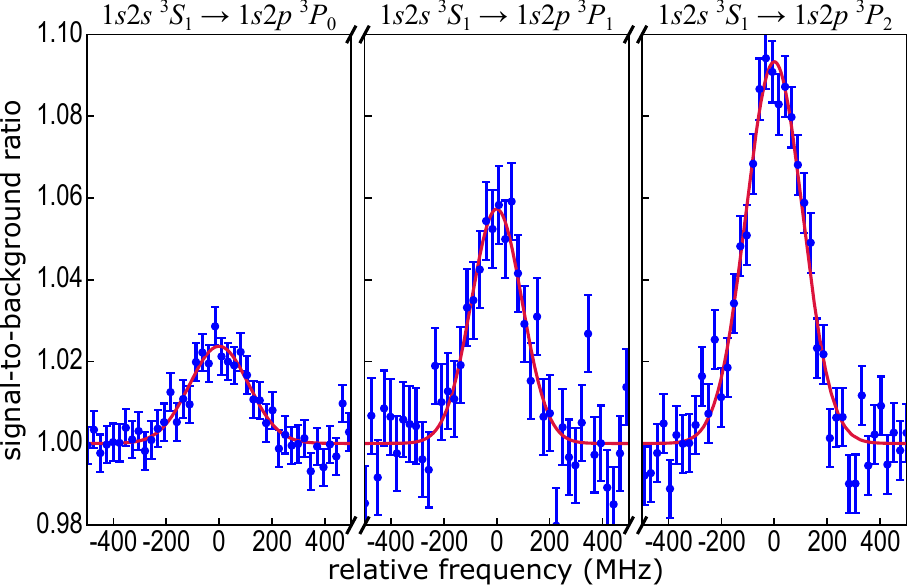}
    \caption{Example spectra of the $1s2s\,^3\!S_1 \rightarrow 1s2p\,^3\!P_J$ transitions in $^{14}$C$^{4+}$. The $x$-axis shows the frequency in the ion rest frame relative to the extracted centroids \replaced{at approximately $227$\,nm}{}. The spectra are normalized to the background and their intensity reflects the relative strengths of the transitions. The measurement time was adjusted in the experiment to achieve a similar signal-to-noise ratio. The data was fitted with a Gaussian, yielding a linewidth of $\sigma\approx 100$\,MHz.} 
    \label{fig:14C-spectrum}
\end{figure}

\begin{figure*}
    \centering
    \includegraphics[width=1\linewidth]{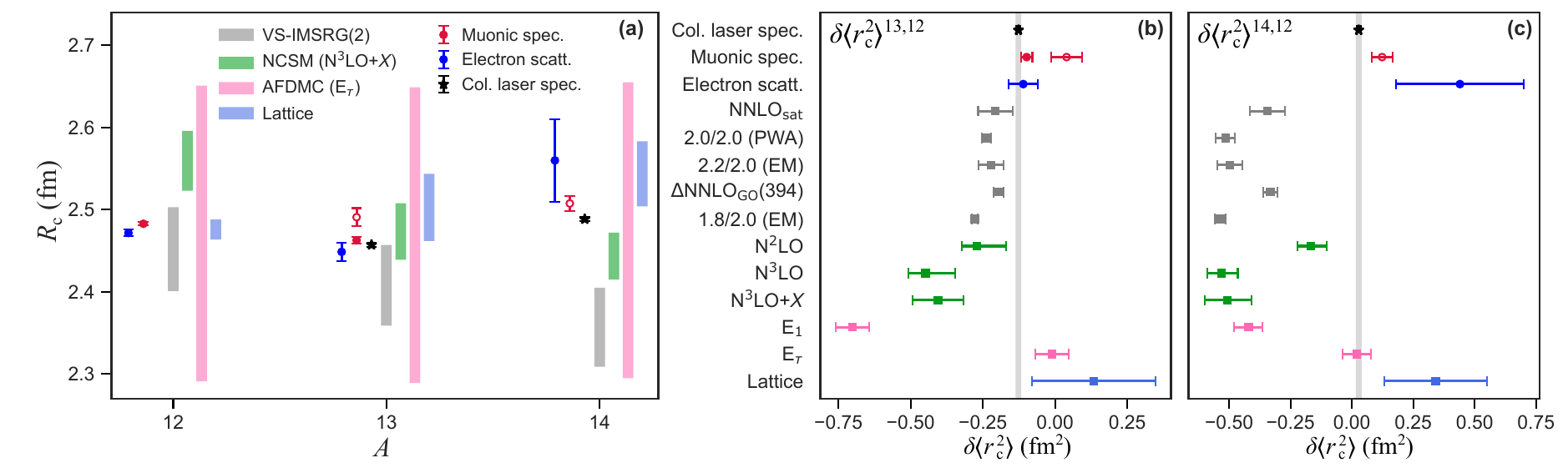}
    \caption{Root-mean-square charge radii $R_\mathrm{c}$ and differential mean-square charge radii $\delta \langle r^2_\mathrm{c}\rangle$ of $^{12,13,14}$C.  Experimental findings from collinear laser spectroscopy (this work, relative to $R_{\mathrm{c},\mu}^{12}$), electron scattering, and muonic spectroscopy are plotted as data points with error bars that depict the full uncertainty. The full circles of the muonic spectroscopy correspond to measurements with a crystal spectrometer~\cite{Ruckstuhl.1984, deBoer.1985} and the open circles to a measurement campaign with a Ge detector~\cite{Schaller.1982}. 
    Theoretical results from VS-IMSRG, NCSM, and AFDMC calculations (this work and \cite{Muller.2025}) as well as nuclear lattice calculations~\cite{Elhatisari.2024} are presented. In (a), they are depicted as bars representing the full uncertainty estimated by order-by-order calculations (NCSM, AFDMC) or the spread between different interactions (VS-IMSRG). The latter show a similar trend but yield different absolute charge radii. In (b) and (c), only the numerical uncertainty is depicted for VS-IMSRG and AFDMC calculations for each Hamiltonian considered, while NCSM and nuclear lattice calculations are given with their full uncertainty.}
    \label{fig:chargeRadii}
\end{figure*}

Laser spectroscopy was performed in parallel and antiparallel geometry quasi-simultaneously with a collinear and anticollinear laser beam, respectively. Two identical continuous-wave laser systems based on Sirah Matisse Ti:Sapphire lasers operated at $\approx 910$\,nm were used. Both were pumped with $20$\,W $532$-nm light from a frequency-doubled Nd:YVO$_4$ laser (Millennia eV, Spectra Physics) and frequency locked to a GPS-referenced frequency comb (Menlo Systems FC1500-250-WG). Each laser's output was frequency quadrupled with two cavity-based frequency doublers (Wavetrain 2, Spectra Physics). About $1$\,mW of the produced UV laser power for each direction was transported to the interaction region, where both beam diameters were adjusted to be about $1.4$\,mm.
The rest-frame frequencies $\nu_0$ of all three fine-structure transitions were determined from collinear (c) and anticollinear (a) spectroscopy by taking the geometric mean and correcting for differences in the applied scan potentials and the photon recoils as described in  \cite{ImgramPRA.2023}.
Spectroscopy was performed in a-c-c-a order to minimize the impact of voltage drifts. The central EBIS electrode potential was actively stabilized to a precision voltage divider~\cite{Koenig.2024}, but the electron beam current and, hence, the effective potential inside the trap, showed slow long-term drifts \cite{ImgramPRA.2023}. Typical spectra of the three fine-structure transitions are depicted in Fig.\,\ref{fig:14C-spectrum}, and the results are listed in Table~\ref{tab:TransitionFreq}. In total, 113 ac pairs were measured for the $1s2s\,^3\!S_1 \rightarrow 1s2p\,^3\!P_2$ transition, yielding a statistical uncertainty of $\Delta \nu_{P_2}=0.53$\,MHz. This was determined as the standard deviation of the mean, which was slightly larger than the uncertainty of the weighted mean ($0.42$\,MHz). For the $1s2s\,^3\!S_1 \rightarrow 1s2p\,^3\!P_1$ transition, 50 pairs ($\Delta \nu_{P_1}=1.4$\,MHz), and the $1s2s\,^3\!S_1 \rightarrow 1s2p\,^3\!P_0$ transition 60 pairs ($\Delta \nu_{P_0}=1.9$\,MHz) were measured. 

\begin{table} [b]
    \renewcommand{\arraystretch}{1.2}
    \centering
    \caption{Experimentally determined transition frequencies in $^{14}$C$^{4+}$. The given total uncertainty consists of the individual statistical uncertainties $<2$\,MHz and a $1.7$-MHz systematic uncertainty. }
    \begin{tabular}{cc}
    \hline
    \hline
         Transition & ~~~~~Frequency (MHz)~~~~~\\
         \hline 
         $1s2s\,^3\!S_1 \rightarrow 1s2p\,^3\!P_0$& 1\,316\,148\,065.8\,\replaced{(27)}{(2.7)}\\
         $1s2s\,^3\!S_1 \rightarrow 1s2p\,^3\!P_1$& 1\,315\,773\,045.2\,\replaced{(23)}{(2.3)}\\
         $1s2s\,^3\!S_1 \rightarrow 1s2p\,^3\!P_2$& 1\,319\,844\,472.9\,\replaced{(18)}{(1.8)}\\
         \hline
    \hline
    \end{tabular}
    \label{tab:TransitionFreq}
\end{table}

A total systematic frequency uncertainty of $1.7$\,MHz was estimated in our previous work~\cite{ImgramPRA.2023}, dominated by a potentially imperfect alignment of the two laser beams in combination with the ion beam divergence, which can lead to an interaction with different velocity classes for both directions. The systematic uncertainty is added in quadrature to the statistical uncertainty, yielding a combined $1\,\sigma$ uncertainty of less than $3$\,MHz.
The fine-structure triplet's center-of-gravity transition frequency was determined according to
\begin{align*}
    \nu_{S\rightarrow P} &= \frac{1}{9} (\nu_{P_0}+3\nu_{P_1}+5\nu_{P_2})\\
    &=1\,318\,076\,618.4\,\replaced{(13)}{(1.3)}\,\mathrm{MHz}
\end{align*}
with a total uncertainty of $1.3$\,MHz. Compared with the result of $^{12}$C~\cite{Imgram.2023}, this yields an isotope shift of $95\,878.9\,\replaced{(17)}{(1.7)}$\,MHz.
The splitting isotope shift of $^{12,13,14}$C was evaluated and compared with NRQED calculations capable of kHz-level accuracy.
Excellent agreement was found, validating the experimental results~\cite{Mueller.2026}.

Employing Eq.\,(\ref{eq:r2}) with $F=-211.5\,(1)$\,MHz/fm$^2$ and $\delta\nu_\mathrm{M} = 95\,884.9\,(5)$\,MHz~\cite{Yerokhin.PrivCom} based on NRQED calculations up to the order $m\alpha^6$~\cite{Yerokhin.2022} yields
\begin{align*}
    \delta \langle r_\mathrm{c}^2 \rangle^{14,12} = 0.028\,(8)\,\mathrm{fm}^2, \\
    \delta R_\mathrm{c}^{14,12} = 0.0056\,(17)\,\mathrm{fm},
\end{align*}
with the root-mean-square charge radius $R_\mathrm{c}^{14}=\sqrt{\left(R_\mathrm{c}^{12}\right)^2+\delta \left\langle r_\mathrm{c}^2 \right\rangle^{14,12}}$. 
To determine $R_\mathrm{c}^{14}$, the reference radius $R_\mathrm{c}^{12}= 2.4829\,(19)$\,fm from muonic spectroscopy~\cite{Ruckstuhl.1984} is employed, which is considered to be the most accurate result since it has been obtained with a crystal spectrometer. This yields $R_\mathrm{c}^{14}= 2.4885(25)$\,fm. 

In Fig.\,\ref{fig:chargeRadii}(a), the root-mean-square charge radii of $^{12,13,14}$C are plotted together with other experimental results from muonic spectroscopy and electron scattering as well as with theoretical predictions. Fig.\,\ref{fig:chargeRadii}(b) and (c) show the differential mean-square charge radii $\delta \langle r_\mathrm{c}^2 \rangle^{13,12}$ and $\delta \langle r_\mathrm{c}^2 \rangle^{14,12}$.
The electron scattering results are the following: $^{12}$C is taken as the weighted average of the values reported in~\cite{Cardman.1980, Reuter.1982, Sick.1982, Offermann.1991}. For $^{13}$C, $\delta \langle r_\mathrm{c}^2 \rangle_\mathrm{el}^{13,12}$ from Ref.\,\cite{Heisenberg.1970} is combined with $R_\mathrm{c,el}(^{12}$C) defined above as this yields a more precise result for $R_\mathrm{c,el}(^{13}$C) than the absolute value reported in~\cite{Heisenberg.1970}. For $^{14}$C, absolute and relative values are taken from~\cite{Kline.1973} as the combination affects neither the absolute value nor its uncertainty. 
While $\delta \langle r_\mathrm{c}^2 \rangle^{13,12}_\mathrm{el}$ is in excellent agreement with the present data, $\delta \langle r_\mathrm{c}^2 \rangle^{14,12}_\mathrm{el}$ is larger but also shows a considerable uncertainty. 

The muonic spectroscopy measurements of $^{12}$C and $^{13}$C, depicted with the full symbol in Fig.\,\ref{fig:chargeRadii}(a), were precisely obtained with a crystal spectrometer and performed consecutively~\cite{Ruckstuhl.1984, deBoer.1985}. For $^{14}$C, only one muonic measurement was performed using a Ge detector for x-ray detection~\cite{Schaller.1982}. In this campaign $^{12,13}$C were also measured. Since their reported values for the relative charge radii are significantly more precise, $R_\mathrm{c,\mu}^{12}$ obtained by the crystal spectrometer is taken as \replaced{the reference value}{ reference}. The combined $R_\mathrm{c,\mu}^{13,14}$ are plotted as open symbols.
While the muonic measurement of $\delta \langle r_\mathrm{c}^2 \rangle^{13,12}$ with the crystal spectrometer (full circles)~\cite{deBoer.1985} agrees with our findings, the Ge-detector-based result (open circles)~\cite{Schaller.1982} deviates by nearly $3\,\sigma$. A similar tension of $1.9\,\sigma$ is observed for $\delta \langle r_\mathrm{c}^2 \rangle^{14,12}$. 
We conclude that the uncertainties of Ref.\,\cite{Schaller.1982} are underestimated, particularly since our results are backed by the excellent agreement with NRQED calculations~\cite{Mueller.2026}.
The accurate determination of $R_\mathrm{c}(^{14}\mathrm{C})$ with proper uncertainty estimation is critical because this radius was established to serve as a reference for the $^{14}$O mirror partner~\cite{Ohayon.2025}.

\begin{figure}
    \centering
    \includegraphics[width=1\linewidth]{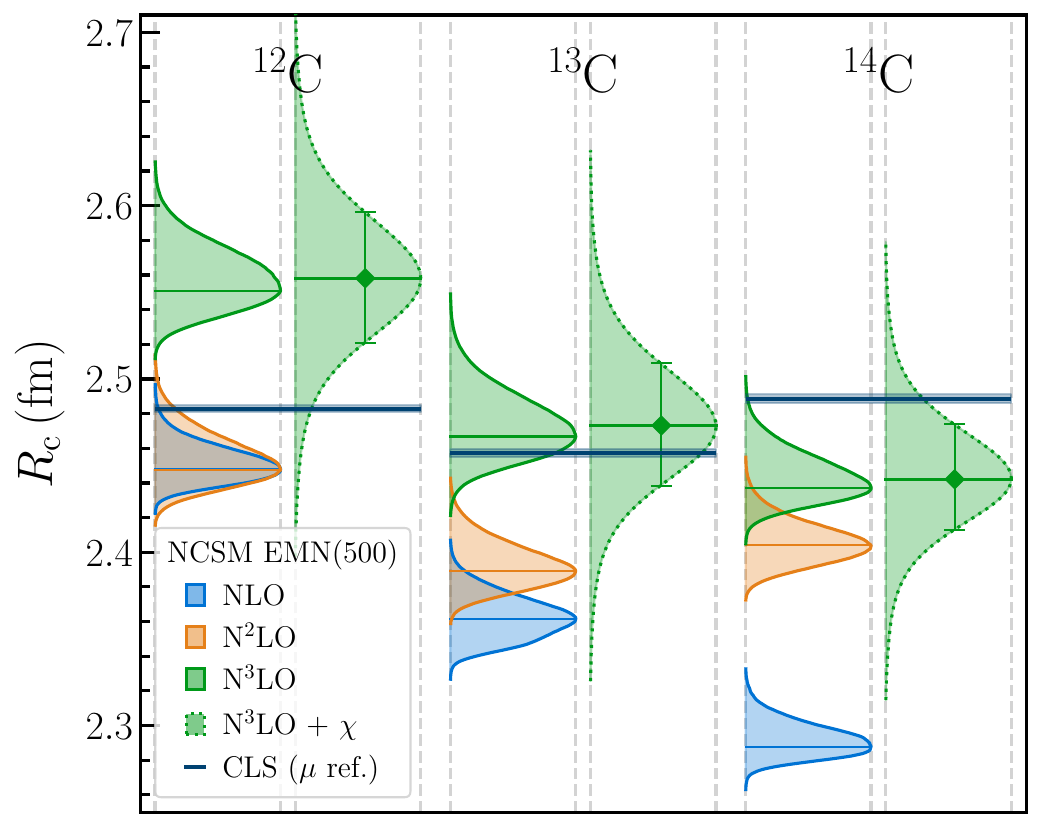}
    \caption{\replaced{NCSM predictions made by ANNs}{ANN predictions} for the charge radii of $^{12,13,14}$C \replaced{}{from NCSM calculations} with non-local EMN Hamiltonians with cutoff $\Lambda = 500$\,MeV.
    These are compared to experimental data from collinear laser spectroscopy (CLS) referenced against the muonic measurement of $^{12}$C. For each isotope, the first column shows the predicted ANN probability distributions up to \nnnlo{}, and the second column combines the \nnnlo{} distribution with EFT truncation uncertainties from a Bayesian analysis.}
    \label{fig:annPred}
\end{figure}

{\it Theory.}
We present \textit{ab initio} computations of the carbon radii based on two- and three-nucleon interactions from chiral effective field theory (EFT)~\cite{Epelbaum:2008ga, Machleidt:2011zz} using complementary many-body methods.
One of the methods is the NCSM, which is based on a direct diagonalization of the Hamiltonian in an $A$-body model space, spanned by harmonic oscillator (HO) single-particle states with oscillator length $a_{\text{HO}}$ and truncated with respect to the total number of HO excitation quanta $N_{\max}$. For the NCSM calculations we employ a family of chiral Hamiltonians up to next-to-next-to-next-to-leading order (\nnnlo{}) in the chiral expansion using the non-local two-nucleon potentials by Entem-Machleidt-Nosyk (EMN)~\cite{EnMa17} with a $500$~MeV cutoff and three-nucleon interactions with consistent chiral order, regulator, and cutoff~\cite{HuVo20}. The Hamiltonians and all other operators are consistently transformed using the similarity renormalization group~\cite{Roth2014evolved} with flow parameter $\alpha=0.08\,\text{fm}^4$ to improve convergence. Our calculations were performed with the code MFDn~\cite{ccpe-10-2013-Aktulga,SHAO20181}.
The NCSM calculations exhibit convergence with increasing $N_\mathrm{max}$ towards an $a_\mathrm{HO}$-independent value representing the exact result. However, for mid-$p$-shell nuclei like carbon, the computational limitations on model-space dimensions and $N_\mathrm{max}$ do not allow for fully converged energies or radii. We have therefore developed a universal machine learning approach based on artificial neural networks (ANN) to provide converged predictions from non-converged NCSM sequences~\cite{KnoWo23,WoKno24,KnoLo25}. The ANNs are trained on calculations with $A \leq 4$, where converged results are available, using sequences of four consecutive $N_{\max}$ (with $N_{\max}\leq 20$) for three $a_\mathrm{HO}$ values as input. A set of $1000$ independently trained ANNs is used, together with the sampling of all available evaluation data, to generate a distribution of predictions that enables a statistical quantification of many-body uncertainties. Details on the training and evaluation process can be found in \cite{WoKno24,WoGe25}. Fig.\,\ref{fig:annPred} shows the probability distributions predicted by the ANNs for the charge radii of $^{12,13,14}$C using interactions from \nlo{} to \nnnlo{} based on NCSM results up to $N_{\max}=8$. The results depend considerably on the chiral order. At \nnnlo{} the radii are larger than at lower orders and show a continuous decrease with increasing neutron number. Interestingly, for \nnlo{} a different radius pattern emerges: The radius decreases from $^{12}$C to $^{13}$C, but then increases from $^{13}$C to $^{14}$C. 
We quantify the EFT truncation uncertainties using a Bayesian approach based on order-by-order convergence~\cite{MeFu19}. Combining EFT and many-body uncertainties~\cite{WoGe25} leads to the distributions shown in the right-hand column of Fig.\,\ref{fig:annPred} for each isotope. 

Furthermore, we compute the charge radii of $^{12-14}$C using the \textit{ab initio} VS-IMSRG~\cite{Tsukiyama:2010rj, Stroberg:2019mxo}  using the VS-IMSRG(2) approximation and a $p$-shell valence space (see End Matter and Ref.~\cite{Muller.2025} for details).
We investigate the Hamiltonian sensitivity of our predictions by considering five Hamiltonians from chiral EFT that differ in their construction and fit to data~\cite{Hebeler:2010xb, Ekstrom:2015rta, Jiang:2020the}
and also quantify model-space uncertainties.
In Fig.\,\ref{fig:chargeRadii}(a), the bands indicate the spread of predictions of the five Hamiltonians, while in Figs.\,\ref{fig:chargeRadii}(b,c) the error bars for each Hamiltonian indicate the model-space uncertainty.
We find that our calculations consistently predict a decreasing trend in charge radii.
Within the range of Hamiltonians, we find that our predictions are compatible with data for $^{12}$C and $^{13}$C, but not $^{14}$C.

We also calculate the charge radii of \(^{12\text{--}14}\)C using the AFDMC method, which employs imaginary-time propagation techniques to obtain the ground state starting from a variational ansatz (see End Matter and Ref.~\cite{Gandolfi:2020pbj} for details). Importantly, AFDMC does not rely on an HO basis expansion, and the variational ansatz already encodes the correct asymptotic behavior of the wave function. Our calculations take as input local chiral EFT Hamiltonians at \nnlo{}~\cite{Gezerlis:2013ipa,Gezerlis:2014zia,Lynn:2015jua} with the harder cutoff $R_0$=1.0 fm.
In Fig.~\ref{fig:chargeRadii}(a), the isospin-dependent three-body contact interaction (E$_\tau$) is shown, and the uncertainty band combines the Monte Carlo statistical error with the uncertainty associated with the order-by-order convergence of chiral EFT, which is about 0.18~fm~\cite{Lonardoni:2017hgs}. As depicted in Figures~\ref{fig:chargeRadii}(b,c), where only the Monte Carlo statistical error is shown, E$_\tau$ yields a very good agreement with experiment. On the other hand, the central three-body contact interaction (E$_1$) yields a charge radius of $^{12}$C that is significantly larger than those of $^{13,14}$C, while still exhibiting an increase from $^{13}$C to $^{14}$C.

{\it Discussion.}
Our measured differential charge radius $\delta\langle r_\mathrm{c}^2 \rangle^{14,12}$ reduces the uncertainty by a factor $30$ compared to electron scattering and by a factor $5$ compared to the best muonic result. Moreover, the deviation from the only muonic measurement of $\delta \langle r_\mathrm{c}^2 \rangle^{14,12}$ from Ref.~\cite{Schaller.1982} suggests its uncertainty may be systematically underestimated, similar to the deviation observed for $^{13}$C when compared to other muonic or laser spectroscopy measurements~\cite{deBoer.1985,Mueller.2024}. This has direct consequences for the analysis of the $^{14}$C--$^{14}$O mirror pair, potentially accessible with highest experimental precision, impacting the extraction of the neutron skin thickness and constraints for the nuclear equation of state~\cite{Brown.2017}.
The improved precision clarifies the charge-radius trend in $^{12,13,14}$C. Unlike lighter nuclei, which are often dominated by cluster-related effects, the carbon isotopic chain provides a clear example of a transition toward the behavior typical of medium-mass and heavier nuclei: The heavier even-even isotope $^{14}$C exhibits a slightly larger charge radius than $^{12}$C, and odd--even staggering is observed, with the odd-$N$ isotope $^{13}$C having a smaller charge radius than its even-$N$ neighbors. Nuclear lattice~\cite{Elhatisari.2017} and antisymmetrized molecular dynamics theory calculations~\cite{Thiamova.2004,Kanada.2015} support weak clustering in the ground state and attribute the trend to a solid proton core.

Comparison with state-of-the-art \textit{ab initio} calculations for $^{12\text{–}14}$C reveals limitations of the present theoretical description. The absolute radii show a large spread across methods, mainly driven by the choice of the nuclear interaction. Currently employed chiral interactions do not provide a very precise description of charge radii in these isotopes. Focusing on the isotopic pattern from $^{12}$C to $^{13,14}$C, most methods and interactions reproduce the decrease from $^{12}$C to $^{13}$C in qualitative agreement with experiment, but generally overestimate its magnitude~\cite{Muller.2025,Mueller.2024}. Similarly, most predict a $^{14}$C radius smaller than both $^{12}$C and $^{13}$C. 
Only the isospin-dependent AFDMC E$_\tau$ results reproduce the experimental trend. Contrarily, the central E$_1$ interaction drastically overestimates the $^{12}$C charge radius but agrees for $\delta \langle r_\mathrm{c}^2 \rangle^{14,13}$. Furthermore, the \nnlo{} interaction used in the NCSM calculations correctly predicts an increasing charge radius from $^{13}$C to $^{14}$C, which is puzzling as the \nnnlo{} interaction from the same family -- which should, in principle, provide a more accurate description -- again predicts a monotonically decreasing trend.
We note that the NNLO$_\mathrm{sat}$ Hamiltonian employed in our VS-IMSRG calculations is optimized to the charge radius of $^{14}$C based on single-reference coupled-cluster calculations~\cite{Ekstrom:2015rta}. Nevertheless, our VS-IMSRG results do not agree with experiment. Calculations for NNLO$_\mathrm{sat}$ using the single-reference IMSRG predict a $0.1$\,fm larger charge radius, compatible with experiment. The sizable difference between valence-space and single-reference IMSRG predictions indicates significant many-body uncertainties and highlights the challenge of describing the carbon isotopes with \textit{ab initio} methods.

The deviating isotopic trend may originate from an overestimation of shell effects in $^{14}$C: Extending VS-IMSRG and AFDMC calculations to more neutron-rich isotopes produces a significant increase of the charge radius to $^{15,16}$C and a kink at $N=8$ (see \cite{data} for additional data). Alternatively, the discrepancy may be due to an underestimation of clustering correlations. Nuclear lattice calculations, which naturally incorporate clustering degrees of freedom, predict an overall increasing trend along the carbon isotopic chain~\cite{Elhatisari.2024}, in contrast to VS-IMSRG and NCSM results. We conclude that currently employed chiral interactions are not yet accurate enough to describe subtle structural effects in charge radii at a level comparable to experimental precision.
A possible way forward is to optimize nuclear Hamiltonians to charge radii in this mass region. Previous efforts optimized Hamiltonians to charge radii of carbon and oxygen isotopes using approximate many-body methods~\cite{Ekstrom:2015rta,HuVo20,Arthuis:2024mnl,Hu:2025cjl}. The new $^{12,14}$C radii can be included in future optimizations, which should account for method uncertainties to avoid overfitting. One attractive option would be a global optimization based on NCSM calculations, enabled by recently developed emulators and extrapolation techniques~\cite{Konig:2019adq,Duguet:2023wuh,WoKno24}. Carbon isotopes are well suited for this purpose, as they are the lightest systems in which NCSM calculations fail to reproduce experimental radius trends while exhibiting a rich interplay of shell and cluster effects.

\section*{Data Availability}
The data that support the findings of this article, including a table with the theoretical results, are openly available~\cite{data}.
   
\section*{End Matter}
\label{sec:SM}

{\it VS-IMSRG calculations.}
The valence-space in-medium similarity renormalization group solves the many-body Schrödinger equation starting from a given intrinsic nuclear Hamiltonian with two- and three-nucleon interactions~\cite{Hergert:2015awm,Stroberg:2019mxo}.
It computes a unitary transformation of the Hamiltonian to decouple a core and valence space from the full Hilbert space.
The remaining valence space problem is then solved via exact diagonalization.
In this work, the unitary transformation $U = e^\Omega$ is truncated at the normal-ordered two-body level, the VS-IMSRG(2) approximation.
We expand our calculations in a basis of 13 major HO shells with the oscillator frequency $\hbar\omega=16\:\mathrm{MeV}$ and start from a Hartree-Fock reference state.
Charge radii are evaluated from the intrinsic point-proton radius squared operator $R_p^2$ including spin-orbit and Darwin-Foldy corrections~\cite{Friar:1997js, Ong:2010gf, Heinz:2024juw} and corrections due to the proton and neutron charge radii squared, $r_p^2=0.7071\:\mathrm{fm}^2$ and $r_n^2=-0.115\:\mathrm{fm}^2$~\cite{Workman.2022}.
For $^{12-14}$C, we employ a $p$-shell valence space, which ends at $^{14}$C with $N=8$.
We explored trends beyond $^{14}$C with additional calculations using a proton $p$-shell, neutron $sd$-shell valence space.
We found that from $^{14}$C to $^{16}$C, the charge radii consistently increase for all Hamiltonians considered.
This yields a pronounced ``kink'' at $^{14}$C, which can be associated with the $N=8$ magic number.

{\it AFDMC calculations.}
The Hamiltonians employed in this work include consistent two- and three-body
potentials derived at N$^2$LO in the chiral
EFT expansion that are local in coordinate space%
~\cite{Gezerlis:2013ipa,Gezerlis:2014zia,Lynn:2015jua}.
These interactions have been extensively validated in AFDMC calculations of
ground-state energies, radii, and magnetic moments of nuclei with up to
$A\simeq 20$ nucleons%
~\cite{Lonardoni:2017hgs,Martin:2023dhl}. For the charge radius of $^{12}$C, the chiral-EFT truncation error has been estimated in Ref.~\cite{Lonardoni:2017hgs} to be 0.18~fm. We assume the same uncertainty for $^{13}$C and $^{14}$C, as it is expected to be similar for neighboring isotopes computed at the same chiral order and with the same regulator.

The AFDMC method~\cite{Schmidt:1999lik,Carlson:2014vla, Gandolfi:2020pbj}
employs imaginary-time projection to extract the ground state of the nuclear
Hamiltonian starting from a suitably chosen variational wave function with
nonvanishing overlap with the exact ground state
\begin{equation}
|\Psi_0\rangle
= \lim_{\tau \to \infty} e^{-(H - E_V)\tau}\,|\Psi_V\rangle \,.
\end{equation}
In the above equation, \(E_V\) is an energy offset introduced to control the normalization of the
propagated wave function and is typically chosen close to the exact ground-state
energy \(E_0\).

The variational wave function is written as the product of a correlation operator
and an antisymmetric mean-field state~\cite{Lonardoni:2017hgs}
\begin{equation}
|\Psi_V\rangle
= \left(F_c + F_{2b} + F_{3b}\right)\,|\Phi\rangle_{J^\pi M T_z} .
\end{equation}
Here, \(F_c\) accounts for spin- and isospin-independent two- and three-body
correlations. To keep the computational cost polynomial in the number of
nucleons \(A\), the operators \(F_{2b}\) and \(F_{3b}\) include linearized
spin- and isospin-dependent two- and three-body correlations, as described in Ref.~\cite{Gandolfi:2014ewa}.

The long-range component \( |\Phi\rangle \) is taken to be a shell-model-like
state with good total angular momentum \(J\), its projection \(M\), parity
\(\pi\), and isospin projection \(T_z\). It is modeled as a linear combination
of Slater determinants,
\begin{equation}
|\Phi\rangle_{J^\pi M T_z}
=
\sum_n c_n
\sum_{K}
C^{J M}_{nK}\,
\mathcal{A}
\big(
\phi_{\alpha_1}^{(n)}\cdots\phi_{\alpha_A}^{(n)}
\big)_K \,,
\end{equation}
where \(\mathcal{A}\) enforces antisymmetrization, and \(K\) denotes the set of
single-particle angular-momentum projections entering determinant \(n\). The
coefficients \(C^{J M}_{nK}\) are Clebsch--Gordan coefficients that couple
those single-particle angular momenta to good total \(J\) and projection \(M\).
The coefficients \(c_n\) are variational parameters multiplying different
components with the same quantum numbers.
The single-particle orbitals \(\phi_\alpha^{(n)}\) are obtained by solving the
single-particle Schr\"odinger equation in a Woods--Saxon potential.

\begin{acknowledgments}
WN acknowledges initiating discussions with U. Köster about the possibility of handling $^{14}$C in a standard laboratory and J. Birkhan for his full support with regard to radiation protection. We acknowledge support by the Deutsche Forschungsgemeinschaft (DFG, German Research Foundation) -- Project-ID 279384907 -- SFB 1245, by the BMFTR under Contract Nos. 05P19RDFN1 and 05P21RDFN1, by the European Research Council (ERC) under the European Union's Horizon 2020 research and innovation programme (Grant Agreement No.~101020842), by the U.S. Department of Energy, Office of Science, under Award No.~DE-SC0023495 (SciDAC5/NUCLEI), and by the Laboratory Directed Research and Development Program of Oak Ridge National Laboratory, managed by UT-Battelle, LLC, for the U.S.\ Department of Energy. 
An award of computer time was provided by the U.S. Department of Energy’s (DOE) Innovative and Novel Computational Impact on Theory and Experiment (INCITE) Program. This research used resources from the Argonne Leadership Computing Facility, a U.S. DOE Office of Science user facility at Argonne National Laboratory, which is supported by the Office of Science of the U.S. DOE under Contract No. DE-AC02-06CH11357.
The work of A.~L.~ is supported by the U.S. Department of Energy, Office of Science, Office of Nuclear Physics, under contract DE-AC02-06CH11357, by the DOE Early Career Research Program, by the Office of Advanced Scientific Computing Research, Scientific Discovery through Advanced Computing (SciDAC) NUCLEI program, and by grant PID2023-147458NB-C21 funded by MCIN/AEI/10.13039/501100011033,             and by the European Union. The work of S.G. is supported by the U.S. Department of Energy through the Los Alamos National Laboratory. Los Alamos National Laboratory is operated by Triad National Security, LLC, for the National Nuclear Security Administration of U.S. Department of Energy (Contract No.~89233218CNA000001), by the Office of Advanced Scientific Computing Research, Scientific Discovery through Advanced Computing (SciDAC) NUCLEI program, and by the LANL LDRD program. The work of T.M. is supported by JST ERATO Grant No. JPMJER2304, Japan, by JSPS KAKENHI Grant Numbers 25K07294, 25K00995, and 25K07330.

This research used resources of the Gauss Centre for Supercomputing e.V. (www.gauss-centre.eu) through the John von Neumann Institute for Computing (NIC) on JUWELS at Jülich Supercomputing Centre (JSC). The authors gratefully acknowledge the computing time provided to them on the high-performance computer Lichtenberg at the NHR Center NHR4CES@TUDa. This is funded by the German Federal Ministry of Education and Research (BMBF) and the Hessian Ministry of Science and Research, Art and Culture (HMWK).
This research used resources provided by the Los Alamos National Laboratory Institutional Computing Program, which is supported by the U.S. Department of Energy National Nuclear Security Administration under Contract No.~89233218CNA000001.
This research used resources of the Oak Ridge Leadership Computing Facility at the Oak Ridge National Laboratory, which is supported by the Advanced Scientific Computing Research programs in the Office of Science of the U.S. Department of Energy under Contract No.~DE-AC05-00OR22725.

\end{acknowledgments}

\bibliography{Literature.bib}

\end{document}